\documentclass[%
twocolumn,
showpacs,
 amsmath,amssymb,
 aps,
prl,
]{revtex4}

\usepackage{graphicx}
\usepackage{dcolumn}
\usepackage{bm}

\begin{document}

\preprint{APS/123-QED}

\title{Heterogeneous diversity of spacers within CRISPR}
\author{Jiankui He$^1$ and Michael W. Deem$^{1,2}$}

\affiliation{\hbox{}$^1$Department of Physics \& Astronomy
and
\hbox{}$^2$Department of Bioengineering, Rice University,
Houston,Texas 77005, USA}

\date{\today}

\begin{abstract}
Clustered regularly interspaced short palindromic repeats (CRISPR) in bacterial and archaeal DNA have recently been shown to be a new type of anti-viral immune system in these organisms.  We here study the diversity of spacers in CRISPR under selective pressure. We propose a population dynamics model that explains the biological observation that the leader-proximal end of CRISPR is more diversified and the leader-distal end of CRISPR is more conserved. This result is shown to be in agreement with recent experiments. 
Our results show that
the CRISPR spacer structure is influenced by and provides a record of the
viral challenges that bacteria face.
\end{abstract}


\maketitle

Clustered regularly interspaced short palindromic repeats (CRISPR) in bacteria and archaea have recently been suggested to provide adaptable immunity in these organisms  \cite{Barrangou2007,Brouns2008,Deveau2008}. A typical CRISPR system is composed of CRISPR-associated (Cas) genes and a CRISPR-cassette \cite{jansen2002,Bolotin2005, pourcel2005}. A CRISPR-cassette is formed by nearly identical repeats of 24--47 bp long nucleotides separated by similar sized, unique spacers. 
Repeats are nearly but not completely palindromic, which leads
to relatively stable RNA secondary structures transcribed from the repeats.
The CRISPR are commonly followed by a conserved AT-rich sequence known as the leader. The leader appears to promote transcription towards the repeats, generating the RNAs that constitute the molecular basis of the CRISPR interference action. Recent studies have proposed that CRISPR and Cas genes function as an anti-viral defense. A considerable fraction of spacer sequences are found to be similar to known phage sequences, that is sequences of viruses which infect bacteria,
indicating that the spacer sequences may derive from phages \cite{Bolotin2005}. Moreover, when bacteria that possess the CRISPR-Cas system are exposed to phage, the surviving individuals appear to have new virus-derived sequences at the leader-proximal end of CRISPR loci \cite{Barrangou2007, Deveau2008}. Further, the acquisition or loss of  CRISPR elements or of Cas protein genes has been directly correlated with phage and plasmid resistance or sensitivity, respectively \cite{Barrangou2007,Deveau2008,Brouns2008}.
 The CRISPR system has begun to attract a
 large amount of attention due to  its immune function and
 its potential role in restricting 
horizontal gene transfer
\cite{Sorek,Marraffini12192008}. Because the CRISPR system directly targets 
nucleotide sequence, it can prevent horizontal gene transfer by
phage transduction, transformation, or conjugation \cite{Marraffini12192008}. 
The CRISPR system also functions as an immune system
to select against phage \cite{horvath2010}, and it is this function
upon which we focus in this Letter.

Recent experiments have demonstrated that the probability
with which different
individuals share the same spacer at the same position varies with
location in the CRISPR system in populations of bacteria and archaea \cite{tyson2008,Andersson05232008}. However, the mechanism by which the phage-bacteria interaction shapes the spacer structure is poorly understood. In this paper, we propose a model that describes why the newly added spacers are more diversified and the old spacers are more conserved due to selective pressure on the CRISPR system. This model expresses an underlying mechanism that shapes the spacer structure. Solution of this model shows that diversity of CRISPR spacers decreases with distance from the leader sequence.

\begin{figure}
\includegraphics[scale=0.42]{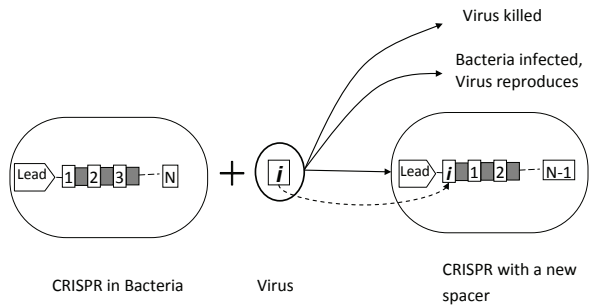}
\caption{\label{cartoon} A schematic representation to describe CRISPR-phage dynamics. Spacers are shown in numbers, and repeats are shown in dark squares. The leader sequence is directly adjacent to the short spacer-repeat units and possibly involved in promoting transcription
towards the repeats. The virus DNA that is recognized by CRISPR is represented by the letter ``i.'' Only the CRISPR of the bacterial genome are shown; other parts of genome are assume to be identical in all bacteria strains. }
\end{figure}

We describe the CRISPR-phage dynamics schematically in Fig.\ \ref{cartoon}.  When bacteria are exposed to phage viruses, there are three possible scenarios: bacteria are infected, viruses are defended, or bacteria acquire new spacers.  In Fig.\ \ref{cartoon}, the bacteria incorporate a piece of the phage DNA represented by the letter `i' into its own genome as a new spacer. New spacers are always added to the leader-proximal end \cite{horvath2010}. To avoid infinite growth of CRISPR, an old spacer is dropped when CRISPR is longer
than a certain length \cite{tyson2008}. The CRISPR system provides an immune response. After insertion of
exogenous DNA from phages or plasmids, the CRISPR spacers are transcribed and processed to CRISPR RNA units. The CRISPR RNA units serve as templates to recognize foreign nucleotide sequence. If any of the CRISPR RNA units match the phage-derived sequences, the phage genetic material is degraded by bacteria. If none of CRISPR RNA units matches the phage-derived sequences, the bacteria are likely to be infected by the phage, and the phages will reproduce. When bacteria divide, the CRISPR are copied to the daughter cells \cite{karginov2010}.

We use a population dynamics model to describe the bacteria-virus community. We assume only one CRISPR locus for each bacteria individual. We first consider a simple case in which there are no more than two spacers for each CRISPR locus. By the first spacer, we mean the spacer that is nearest to the leader sequence. The second spacer is the spacer that is the next nearest to the leader sequence. We consider the following system of ordinary differential equations:
\begin{eqnarray}
\frac{dx_{i,j}}{dt}=cx_{i,j}-\beta\sum_ {k\neq i,j}v_k x_{i,j} + \beta \gamma \sum_{m} x_{j,m} v_i  \\
\frac{dv_k}{dt}=r v_k - \beta \sum_{i,j}x_{i,j} v_k(\delta_{i,k}+\delta_{j,k})
\label{eq}
\end{eqnarray}
 There are two variables in the above equations: $v_k$ is the population of virus strain $k$, and $x_{i,j}$ is the population of bacteria with CRISPR with spacers $i$ and $j$. The first spacer recognizes virus strain $i$ and the second spacer recognizes virus strain $j$. In the absence of phage infection, the bacterial growth is exponential at rate $c$. The term $\beta\sum_ {k\neq i,j}v_k x_{i,j}$ represents the bacteria with spacers of type $i$ and $j$ infected by viruses strains other than $i$ or $j$. Bacteria can be infected or killed when they are exposed to viruses that bacteria do not recognize by CRISPR. The exposure rate of bacteria to virus is $\beta$. The term $\beta \gamma \sum_{m} x_{j,m} v_i $ represents the process of the converting other types of bacteria into bacteria of type $i,j$.  When bacteria of type $j,m$ incorporate virus of strain $i$ into their own genome and add a new spacer, bacteria type $j,m$ are converted to type $i,j$. The probability of adding a new spacer when a bacteria is exposed to a virus is $\gamma$. In the absence of resistance from CRISPR, viral growth is exponential at rate $r$. The term $\beta \sum_{i,j}x_{i,j} v_k(\delta_{i,k}+\delta_{j,k})$ represents the degradation of viruses by bacteria. If any spacers of bacteria of type $i,j$ match viruses of  strain $k$, the bacteria degrade the viruses. The Kronecker delta function $\delta_{i,k} $ is 1 if spacer type $i$ matches virus strain $k$; otherwise, it is 0.
 This model is modified from the classic immune response model with antigenic variation \cite{nowak2000}. In this model, we take only the essential factors into consideration. We do not distinguish the lysis and lysogeny cycle.  Horizontal gene transfer is not considered. Furthermore, because viruses usually have more than one type of host to infect, viral growth is not limited by the
abundance of one specific type of target bacteria \cite{nowak2000,Andersson05232008}.

\begin{figure}
\includegraphics[scale=0.3]{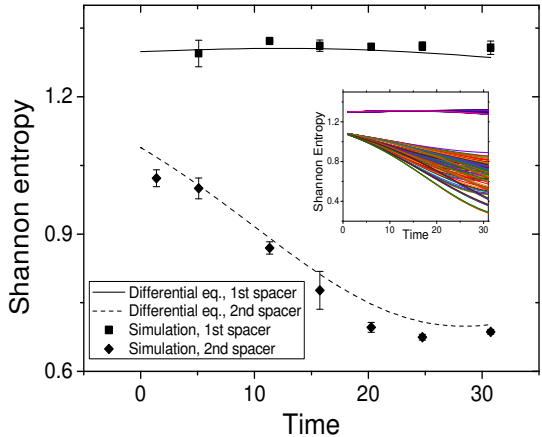}
\caption{\label{twospacer} Diversity of two spacers of CRISPR with time. The differential equation solution and simulation are based on the parameter values $c=0.15, \beta=2\times10^{-6}, \gamma=0.1$, and $r=0.01$. The viruses have four strains (length of string $n=2$) with an initial population ratio 6:2:1:1.  
In the stochastic simulation, the maximal population size is $10^6$ for virus and  $10^5$ for bacteria. Diversity is measured by Shannon entropy. Other measures of diversity such as Simpson's index of diversity give similar results. Error bars are one standard error. The insert figure is solutions of differential equations with 200 different parameter combinations using  Latin hypercube sampling.
The parameter ranges we used are: $c\in(0.01,0.15), \beta\in (10^{-5},2\times10^{-5}), \gamma\in(0.01,0.1), r\in(0.01,0.1) $. We used 200 samplings of this parameter space. 
 The up branches are the first spacer, and the down branches are the second spacers. }
\end{figure}

 Solution of the model shows that the diversity of the old spacer decreases upon challenge by viruses. We solve the differential equations by a standard numerical Runge-Kutta method. The initial value for the differential equations are naive bacteria whose CRISPR provide no resistance to viruses because their spacers are empty. The population of bacteria initially drops rapidly. Some bacteria acquire spacers from viruses and therefore develop resistance. By this means,
the population of bacteria is steadily recovered. We measure the diversity of spacers by the Shannon entropy:
\begin{eqnarray}
D_1=-\sum_i(\sum_j P_{i,j}) \ln(\sum_j P_{i,j})\\
D_2=-\sum_j(\sum_i P_{i,j}) \ln(\sum_i P_{i,j})\\
P_{i,j}=\frac{x_{i,j}}{\sum_{m,n}x_{m,n}}
\label{eq2}
\end{eqnarray}
Here, $D_1$ and $D_2$ are the diversity for the first and second spacers. Because new spacers are always added to the leader-proximal end, the first spacer is ``younger" than the second spacer. If there is no selective pressure on CRISPR, or CRISPR do not provide resistance against viruses, the diversity of spacers along CRISPR should be homogeneous, $D_1=D_2$, because adding and deleting spacers is completely random. With the selective pressure on CRISPR to evolve resistance to phage, we observe a decline of diversity of the second spacer, as shown in Fig.\ \ref{twospacer}.
At the beginning, both positions have high diversity of spacers. With the continuous challenge of viruses and selective pressure for the effective resistance against viruses, the diversity of spacers at the second position decreases with time. When steady state is reached after some time, we observe that the diversity of spacers at the second position is lower than that at the first spacer.  

Our observation  is true for a broad choice of parameters.
The parameter space was explored by using the statistical technique of Latin hypercube sampling (LHS). LHS selects combinations of parameter values from parameter value range and probability distribution function. 
In the insert of Fig.\ \ref{twospacer}, we observe that diversity of the old spacer is decreasing and the diversity of the young spacer is nearly constant over time for all samplings.

Selection for bacteria that contain the most effective spacers decreases the diversity of the old spacer. The bacteria randomly take virus genomes from the environment and incorporate a corresponding spacer. Therefore, the diversity of the first spacer approaches the diversity of viruses in the environment. If the spacers match the dominant virus strain, bacteria containing these spacers are more likely to survive, and therefore spacers that match dominant viruses accumulate in the CRISPR. Bacteria that contain unused spacer elements that provide little protective potency are more likely to be infected by phage.  The spacers corresponding to the dominant virus strain are enhanced and accumulate at the second spacer position.   In other words, if neither the first nor second spacer
matches the dominant viral strain, the bacteria is likely to be eliminated.
If only the first spacer matches the dominant strain, after the next 
spacer incorporation, the matching spacer becomes the second spacer. 
For these two reasons, the diversity of the second spacer is lower than that
of the first.  The diversity of the second spacer is a function of
the viral population diversity and the fitness pressure of the viruses upon
the bacteria.

We seek to identify finite size effects by a stochastic simulation. 
We use the
Lebowitz-Gillespie algorithm to sample the Markov process
with the rates as described in Eqs.\ (1--2).
for the bacteria and virus populations.
Each bacteria and each virus is individually tracked. Viruses are represented as bit-strings. Each bit has two alleles, designated as a ``1'' or ``0.'' In 
the simulation, the length of virus strings are $n$; therefore $2^n$ genotypes are available for viruses. For bacteria, we the consider CRISPR locus only. Each spacer is $n$ bits long \cite{karginov2010}, the same size as viruses. The simulation starts with a population of viruses of different genotypes and bacteria without spacers in CRISPR locus. Viruses infect bacteria with a contact rate $\beta$. If any spacer of a bacterium matches the infecting virus, the virus is killed. Otherwise, the bacterium is infected and dies. Bacteria and viruses reproduce at rate $c$ and $r$ respectively.  Bacteria add a new spacer with a rate $\gamma$ from contacting virus. We show in Fig.\ \ref{twospacer} the simulation
results falling along the infinite-population, mean-field results from
solving the differential equations (1--2). 

\begin{figure}
\includegraphics[scale=0.25]{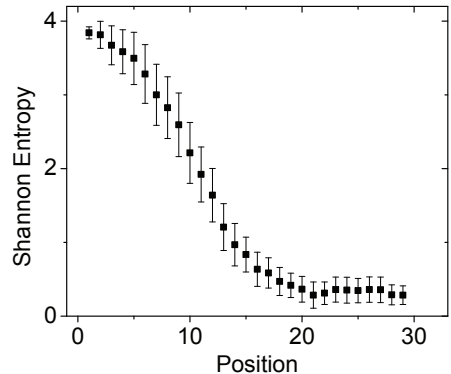}
\caption{\label{multiplespacer} Diversity of spacers at different positions of CRISPR, when the system reaches steady state. The positions with a small number in the $x$-axis are leader-proximal.  In this extended simulation, we use the parameters: $c=0.15, \beta=2\times10^{-5}, \gamma=0.1$, $ r=0.05$, mutation rate per sequence of $\varepsilon=0.01$, size of virus bit-string $n=10$. Initially, there are 150 phage strains with a logarithmic population distribution \cite{tyson2008}. Other parameter settings give similar results. Error bars are one standard error.}
\end{figure}

We further extend our individual-based simulation  to allow  the CRISPR to  have more spacers, random loss of spacers, and mutation. Most CRISPR contain fewer than 50 repeat-spacer units. For example, the average number of spacers of \textit{Streptococcus thermophilus} is 23 per CRISPR locus in one study \cite{Horvath12072007}. In our extended simulation, when the array of spacers of bacteria is longer than 30, a spacer is randomly deleted with probability proportional to its distance to the leader sequence. When viruses replicate, the mutation rate per sequence is $\varepsilon$. We perform mutation by randomly flipping one bit of the virus's bit-string from ``1'' to ``0'' or from ``0'' to ``1.''
 This extended simulation starts with a population of 150 virus genotypes and bacteria without spacers.
 The simulation runs until it reaches steady state. We run the simulation 100 times and average the results. After the simulation reaches steady state, we calculate the diversity of spacers for each position by Shannon entropy.  In Fig.\ \ref{multiplespacer}, we observe that the ``young'' spacers which are leader-proximal are highly diversified and that the ``old'' spacers which are leader-distal are more conserved.

These results support the following scenario:
Infection by a novel viral genotype results in the lysis or weakening of most individuals, except those that are able to capture and incorporate a corresponding spacer into their CRISPR locus. Resistant individuals  rapidly gain a selective advantage, leading to the fixation of the resistant spacer. Increasing polymorphism toward the leader-proximal end provides support that the CRISPR are an actively evolving and functioning phage defense mechanism.

\begin{figure}
\includegraphics[scale=0.25]{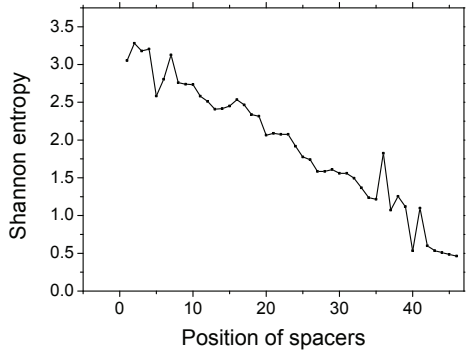}
\caption{\label{horvath_fig} Diversity of spacers of CRISPR loci 1 of \textit{S. thermophilus} strains \cite{Horvath12072007}. The positions with a small number in the $x$-axis are leader-proximal.}
\end{figure}

This model is in agreement with recent experiment results. Horvath \emph{et al.}\ \cite{Horvath12072007} sequenced the CRISPR regions of 124 \textit{S. thermophilus} strains and analyzed 3626 spacers, 926 of which are unique. We aligned the spacers of CRISPR loci 1 for 124 strains. The Shannon entropy was calculated for each aligned position, see Fig.\ \ref{horvath_fig}. Spacers at leader-proximal positions are more diverse and spacers at leader-distal positions are  highly conserved across strains. For example, at the most leader-distal position, 34 of 124 strains share the identical spacer.

\begin{figure}
\includegraphics[scale=0.25]{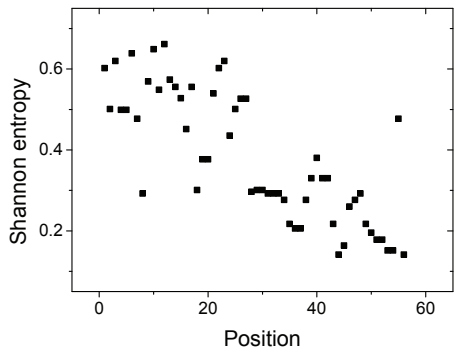}
\caption{\label{experiment2_fig} Diversity of spacers of CRISPR loci of \emph{Leptospirillum} species.  The data are noisy because the CRISPR loci sequence data of \emph{Leptospirillum} are fragmented.}
\end{figure}
 Recent metagenomic studies of environmental microbial samples
provide a population-wide view of the dynamics between phage and CRISPR of the hosts \cite{tyson2008,Andersson05232008,heidelberg2009}.  In one study,
sequence data were assembled from biofilm community samples \cite{tyson2008,Andersson05232008}.
The CRISPR loci of the predominant \emph{Leptospirillum} species display extensive polymorphism. We calculate the Shannon entropy for each position of CRISPR, see Fig.\ \ref{experiment2_fig}.
The bacteria community shared spacer sequences at the leader-distal end of their CRISPR loci, while the leader-proximal end of the loci contained spacers that were mostly unique to each individuals.
The decrease of diversity of spacers from leader-proximal end to leader-distal end supports
a model in which highly plastic CRISPR loci continuously respond to challenge from a rapidly evolving pool of phage.

In summary, the CRISPR system provides adaptable immunity to bacteria and archaea. Bacteria continuously incorporate nucleotide material from phage genomes into CRISPR to gain resistance against phage infection. Viruses continuously perform nucleotide mutation and recombination to avoid being recognized. The coevolution interaction between viruses and the bacteria CRISPR system has shaped the spacer structure of the CRISPR locus. Recent experiments show a decline of diversity of spacers towards the leader-distal end, which our model suggests is a result of selection for the anti-phage protection conferred by the spacers and implies that the CRISPR is an active anti-viral system. That is, an underlying mechanism to shape the spacer structure is the selection of bacteria CRISPR systems that best match with viruses in the environment. 
CRISPR spacer structure is influenced by and provides a record of the
viral challenges that bacteria face.

\bibliography{jiankui}

\end{document}